\documentclass[prd,amsmath,amssymb,nofootinbib,superscriptaddress,twocolumn,10pt]{revtex4-1}

\pdfoutput=1

\usepackage{CJKutf8}
\usepackage{graphicx}
\usepackage{amsmath}
\usepackage{subfigure}
\usepackage{siunitx}
\usepackage{amssymb}
\usepackage{float}
\usepackage[colorlinks=true, linkcolor=red, citecolor=blue]{hyperref}
\usepackage{booktabs}
\usepackage{setspace}
\usepackage{etoolbox}
\AtBeginEnvironment{thebibliography}{\setstretch{1.3}}  % 1.3 倍行距

\newcommand{\orcid}[1]{%
  \href{https://orcid.org/#1}{%
    \includegraphics[height=2ex, keepaspectratio]{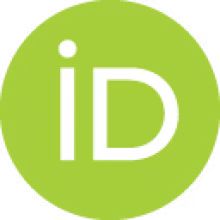}%
  }%
}

\begin{document}

\title{Thermodynamic geometry as the missing link: toward a unified framework for black hole first-order phase transitions}

\author{Shi-Hao Zhang\orcid{0009-0009-2439-508X}}
%\email{pumazhang200@163.com}
%ORCID: 0009-0009-2439-508X
\affiliation{Liaoning Key Laboratory of Cosmology and Astrophysics, College of Sciences, Northeastern University, Shenyang 110819, China}

\author{Jing-Fei Zhang\orcid{0000-0002-3512-2804}}
%ORCID: 0000-0002-3512-2804
%\email{jfzhang@mail.neu.edu.cn}
\affiliation{Liaoning Key Laboratory of Cosmology and Astrophysics, College of Sciences, Northeastern University, Shenyang 110819, China}
\author{Xin Zhang\orcid{0000-0002-6029-1933}}
%ORCID: 0000-0002-6029-1933
\thanks{Corresponding author}
\email{zhangxin@mail.neu.edu.cn}
\affiliation{Liaoning Key Laboratory of Cosmology and Astrophysics, College of Sciences, Northeastern University, Shenyang 110819, China}
\affiliation{MOE Key Laboratory of Data Analytics and Optimization for Smart Industry, Northeastern University, Shenyang 110819, China}
\affiliation{National Frontiers Science Center for Industrial Intelligence and Systems Optimization, Northeastern University, Shenyang 110819, China}

%\date{\today}

\begin{abstract} 
%Black hole first-order phase transitions have been described by several frameworks that originally developed independently, including local geometry, global topology, complex analysis, and thermodynamic geometry. Although the first three have recently been unified, thermodynamic geometry has remained outside this unified picture. We establish a precise connection between the normalized Ruppeiner curvature scalar $R_N$ and the temperature function $T(r_h)$: we prove that the divergence points of $R_N$ coincide exactly with the solutions of $T'(r_h)=0$, where $r_h$ is the horizon radius. These solutions include both extremal points (spinodal points) and stationary inflection points (thermodynamic critical points). Therefore, the divergence of $R_N$ is a necessary but not sufficient condition for a first-order phase transition. This result clarifies the underlying mathematical structure of curvature divergence that has long remained unclear, and explains why thermodynamic geometry can reliably indicate phase transitions but cannot alone confirm them. Using the local geometric framework as a central framework, we incorporate Ruppeiner geometry into this unified framework. A similar analysis applies to Weinhold geometry. Consequently, the four frameworks are unified within a single mathematical structure rooted in the local folding of the temperature function. This unification advances our understanding of the mathematical structure of black hole first-order phase transitions and provides clues for possible extensions to other types of phase transitions.

Black hole first-order phase transitions have been described by several seemingly independent frameworks, including local geometry, global topology, complex analysis, and thermodynamic geometry. While the first three have been unified, thermodynamic geometry has remained outside. We prove that the divergence points of the normalized Ruppeiner curvature scalar $R_N$ coincide exactly with the solutions of $T'(r_h)=0$, where $r_h$ is the horizon radius. These solutions include extremal points (spinodal points) and stationary inflection points (thermodynamic critical points). Thus, the divergence of $R_N$ is a necessary but not sufficient condition for a first-order phase transition. This clarifies the mathematical origin of curvature divergence and why thermodynamic geometry can reliably indicate but not alone confirm phase transitions. Using the local geometric framework as a central framework, we incorporate Ruppeiner geometry into this unified picture; a similar analysis also applies to Weinhold geometry. Consequently, the four frameworks are unified within a single structure based on the local folding of the temperature function. This advances our understanding of the mathematical structure of black hole first-order phase transitions and provides clues for possible extensions to other types of phase transitions.

\end{abstract}
	

\maketitle

\section{Introduction}\label{chap:1}
The small/large black hole first-order phase transition in charged AdS black holes, analogous to the van der Waals fluid, has attracted widespread attention since its discovery \cite{Chamblin:1999tk,Kubiznak:2012wp,Kubiznak:2016qmn,Kubiznak:2014zwa,Mann:2025xrb}. Around this phenomenon, many effective research frameworks have emerged, such as the global topological framework \cite{Wei:2022dzw,Wei:2021vdx,Wei:2024gfz,Wu:2024rmv,Wei:2024gfz,Wu:2024asq,Wu:2024txe,Ai:2025vno,Wu:2025xxo,Yang:2025uul,Wei:2026upy}, the complex analysis framework \cite{Xu:2023vyj,Xu:2025jrk,Li:2025lrq,Guo:2026xlk}, the local geometric framework \cite{Zhang:2025inl}, and thermodynamic geometry \cite{Weinhold:1975xej,Weinhold:1975fyh,Ruppeiner:1979bcp,Ruppeiner:1995zz}. These frameworks provide perspectives beyond traditional thermodynamic analysis: instead of directly computing and comparing free energies, they characterize phase transitions through different mathematical features, such as the topological winding numbers, Riemann surface foliation numbers, extremal points of the temperature function, or curvature scalar divergence. Their value lies in providing different diagnostic tools and in enhancing the understanding of the structure of first-order phase transitions from different mathematical perspectives.

Among them, for the study of the mathematical origin behind black hole first-order phase transitions, the local geometric framework based on the Maxwell equal-area law not only successfully explains the multivalued behavior of physical or geometric quantities (such as the Lyapunov exponent \cite{Guo:2022,Yang:2023,Lyu:2024,Kumara:2024,Du:2025,Shukla:2024,Gogoi:2024,Chen:2025xqc,R:2025gok,Awal:2025irl,Yang:2025,Kumar:2025kzt,Guo:2025pit,Bezboruah:2025udi,Ali:2025ooh,Xie:2025auj,MalikSultan:2026pok,Becar:2026epq,Cheng:2026dnd}, photon sphere radius \cite{PS01,PS02,PS03,PS04,PS05,PS06,PS07,PS08,PS09}, and curvature \cite{Zhang:2025cdx,Zhang:2025kqd}) as functions of temperature in first-order phase transitions found in previous studies, but has also been used to quickly determine whether a black hole has a first-order phase transition and to classify black holes. Recent work has further proved that the local geometric framework, the global topological framework, and the complex analysis framework have exact mappings in the real domain, forming a unified framework \cite{Zhang:2026sht}.

Independent of the above three frameworks, thermodynamic geometry is a
nother widely used analytical framework. It constructs thermodynamic metrics (such as the Ruppeiner metric and the Weinhold metric) and examines whether their curvature scalars exhibit singular behavior to indicate whether a black hole may have a first-order phase transition \cite{Cai:1998ep,Aman:2003ug,Aman:2005xk,Wei:2010yw,Wei:2012ui,Ruppeiner:2013yca,Mansoori:2013pna,Zhang:2014uoa,Zhang:2015ova,Xu:2020gud,Wei:2019yvs,Wei:2020poh,Wu:2020fij,Xu:2020ngu,Wang:2021vbn,Guo:2021wcf,Wang:2022ska,Wang:2022bqu}.

This raises two key questions. First, if thermodynamic geometry describes the same physical process as the local geometric, global topological, and complex analysis frameworks, does it also admit a connection to these three frameworks? Does a larger unified framework exist? Second, although previous studies hold that the divergence of the curvature scalar carries information about black hole first-order phase transitions, the underlying mathematical reason has not been clarified: why can thermodynamic geometry reliably study first-order phase transitions?

In this paper, using the local geometric framework, we prove that the divergence points of the normalized curvature scalar coincide with the solution set of $T'(r_h)=0$ for the temperature function $T(r_h)$, where $r_h$ is the horizon radius. These solutions include both extremal points (spinodal points) and stationary inflection points (thermodynamic critical points). This equivalence clarifies the mathematical origin of curvature divergence: it is a necessary condition for a first-order phase transition, but cannot alone confirm the existence of such a phase transition. Taking the local geometric framework as a central framework, we unify thermodynamic geometry (Ruppeiner and Weinhold geometries) with the other three frameworks within a single mathematical framework, and reveal the mathematical origin underlying curvature scalar divergence and its precise connection to first-order phase transitions. This result indicates that the study of black hole first-order phase transitions is forming a complete unified picture.

The content of this paper is organized as follows: Sec.~\ref{chap:2} reviews the local geometric framework, Sec.~\ref{chap:3} reviews thermodynamic geometry (taking Ruppeiner geometry as an example), Sec.~\ref{chap:4} establishes the connection between the curvature scalar and the temperature function in first-order phase transitions, Sec.~\ref{chap:5} discusses the programmatic significance of the unified framework, and Sec.~\ref{chap:6} presents the summary and outlook. We set $G=c=k_B=\hbar=1$ in this work.

\section{Review of the local geometric criterion} \label{chap:2}
The local geometric criterion is a criterion for small/large black hole first-order phase transitions based on the Maxwell equal-area law and abstracted from the local behavior of the temperature function. In ensembles where the Maxwell equal-area law holds, for any black hole, provided that $T(r_h)$ is at least twice continuously differentiable and not a constant function, the black hole possesses a first-order phase transition if and only if
\begin{align}
\mathrm{card}\{r_h > r_m \mid T'(r_h) = 0\} \geq 2,
\end{align}
where $r_m$ is the minimum horizon radius of the black hole, and the notation ``card'' denotes the cardinality of a set (for example, $\mathrm{card}\,\varnothing = 0$ for the empty set). For a finite set, its cardinality is the number of elements it contains. Therefore, the local criterion can be stated as follows: within the parameter space allowed by physical constraints, when the temperature function of a black hole has two or more extremal points, the black hole possesses a small/large black hole first-order phase transition \cite{Zhang:2025inl}.

This criterion successfully captures the sudden folding behavior of the black hole parameter space at the thermodynamic critical point. Since the black hole first-order phase transition is essentially a cusp catastrophe \cite{Zhang:2025inl,Zhang:2026catastrophe}, the local geometric criterion is in fact a concrete manifestation of a universal law of the cusp catastrophe. This allows us to classify black holes according to the number of extremal points of their temperature functions.

Interestingly, this framework has been proved to have exact mappings in the real domain with two other frameworks, namely the global topological framework and the complex analysis framework, both of which can distinguish whether a black hole possesses a first-order phase transition \cite{Zhang:2026sht}, and the relevant dictionaries have been studied in modified gravity \cite{Hao:2026cco}. Since the different characteristic information about black hole first-order phase transitions obtained from different frameworks is essentially different projections of the cusp catastrophe in different theories, it is reasonable to speculate whether the characteristic behavior found in the thermodynamic geometry of black holes can also be traced back to the characteristic changes of the local behavior of the black hole temperature function described by the local geometric framework.

In the following, we will focus on revealing that the characteristic behavior of Ruppeiner geometry in the thermodynamic geometry of black holes regarding first-order phase transitions is essentially the projection of the local geometric criterion in thermodynamic geometry. A similar analysis also applies to Weinhold geometry.

\section{Review of Ruppeiner geometry}\label{chap:3}
Ruppeiner geometry is a thermodynamic metric/geometry constructed from the Hessian matrix of entropy. Its line element is given by
\begin{align}
\Delta l^2 = -\frac{\partial^2 S}{\partial x^\mu \partial x^\nu} \Delta x^\mu \Delta x^\nu = g_{\mu\nu}^{(R)} \Delta x^\mu \Delta x^\nu,
\end{align}
where $S$ is the entropy of the black hole, $x^\mu$ are thermodynamic variables, and $g_{\mu\nu}^{(R)}$ is the Ruppeiner metric. $\Delta l^2$ measures the distance between two neighbouring fluctuation states, so Ruppeiner geometry has a clear physical meaning \cite{Ruppeiner:1995zz}. We choose $(T,\,V)$ as the thermodynamic variables (fluctuation coordinates), where $V$ is the thermodynamic volume. Then the line element of Ruppeiner geometry is
\begin{align}
dl^2 = \frac{C_V}{T^2} dT^2 - \frac{(\partial_V P)_T}{T} dV^2,
\end{align}
where the heat capacity at constant volume is $C_V=T(\partial_T S)_V$. The corresponding curvature scalar (Gaussian curvature) is
\begin{widetext}
\begin{equation}
    \begin{split}
    &R = \frac{1}{2C_V^2 (\partial_V P)^2} \Bigl\{ T(\partial_V P) \bigl[ (\partial_T C_V)(\partial_V P - T\partial_{T,V} P) + (\partial_V C_V)^2 \bigr] \\
    &+ C_V \bigl[ (\partial_V P)^2 + T \bigl( (\partial_V C_V)(\partial_V^2 P) - T(\partial_{T,V} P)^2 \bigr) + 2T(\partial_V P) \bigl( T(\partial_{T,T,V} P) - (\partial_V^2 C_V) \bigr) \bigr] \Bigr\}.\label{R}
    \end{split}
\end{equation}
\end{widetext}
However, that constant volume means the volume is fixed, i.e., $dr_h=0$. Since the entropy is $S=\pi r_h^2$, this leads to $dS=0$, so that $C_V=0$ holds identically, which makes the thermodynamic metric non-invertible. We can avoid this problem by considering the limit $C_V \to 0^+$ and defining a normalized scalar curvature \cite{Wei:2019yvs,Wang:2021vbn}
{\small
\begin{align}
R_N = C_V R = \frac{(\partial_V P)^2 - T(\partial_{T,V} P)^2 + 2T^2 (\partial_V P)(\partial_{V,T,T} P)}{2(\partial_V P)^2}.\label{RN}
\end{align}}
The divergence of $R_N$ is generally believed to be related to first-order phase transitions \cite{Cai:1998ep,Aman:2003ug,Aman:2005xk,Wei:2010yw,Wei:2012ui,Ruppeiner:2013yca,Mansoori:2013pna,Zhang:2014uoa,Zhang:2015ova,Xu:2020gud,Wei:2019yvs,Wei:2020poh,Wu:2020fij,Xu:2020ngu,Wang:2021vbn,Guo:2021wcf,Wang:2022ska,Wang:2022bqu}, and it is a necessary condition for a black hole to possess a first-order phase transition. Note that $R_N$ is a scalar probe that eliminates the degeneracy problem, and its divergence points are determined by the singularities of the original $R$.

In the following, we will prove that the divergence points of $R_N$ for black holes actually correspond to the extremal points (spinodal points) and stationary inflection points (thermodynamic critical points) of the temperature function $T(r_h)$.

\section{Curvature Scalar and Temperature Function}\label{chap:4}
In this section, we will show how to universally connect the divergence points of $R_N$ with the extremal points and stationary inflection points of $T(r_h)$, and take the Reissner-Nordstr\"om--anti-de Sitter (RN-AdS) black hole as a concrete example.
\subsection{$R_N\to \infty$ and $T'(r_h) = 0$}
Consider the equation of state of an arbitrary black hole: $F(P,V,T)=0$, where $P$ is the pressure, $V$ is the volume, and $T$ is the temperature. From the complete differential of the equation of state
\begin{align}
dF = \frac{\partial F}{\partial P} dP + \frac{\partial F}{\partial V} dV + \frac{\partial F}{\partial T} dT = 0,
\end{align}
we obtain the cyclic relation
\begin{align}
\left( \frac{\partial T}{\partial V} \right)_P \left( \frac{\partial P}{\partial T} \right)_V \left( \frac{\partial V}{\partial P} \right)_T = -1.
\end{align}
This leads to
\begin{align}
\left( \frac{\partial T}{\partial V} \right)_P = -\frac{(\partial_V P)_T}{(\partial_T P)_V}.\label{TV}
\end{align}
From the definition of the curvature scalar $R_N$ in Eq.~(\ref{RN}), we know that for $R_N$ to diverge, it is required that $(\partial_V P)_T=0$. From Eq.~(\ref{TV}), as long as $(\partial_T P)_V$ is finite and nonzero, $(\partial_V P)_T=0$ leads to $\left( \frac{\partial T}{\partial V} \right)_P=0$. In black hole thermodynamics, $V=\frac{4}{3}\pi r_h^3$, which means
\begin{align}
\left( \frac{\partial T}{\partial V} \right)_P = \left( \frac{\partial T}{\partial r_h} \right)_P \left( \frac{\partial r_h}{\partial V} \right)_P.
\end{align}
Since $(\partial r_h/\partial V)_P = [(\partial V/\partial r_h)_P]^{-1}$, for most black holes, as long as $V$ is a monotonic function of $r_h$ ($\left( \frac{\partial r_h}{\partial V} \right)_P \neq 0$), then $\left( \frac{\partial T}{\partial V} \right)_P=0$ means $\left( \frac{\partial T}{\partial r_h} \right)_P=0$. This means that the points at which $R_N$ diverges are either spinodal points ($\frac{\partial T}{\partial r_h} = 0,\,\frac{\partial^2 T}{\partial {r_h}^2} \neq 0$) or thermodynamic critical points ($\frac{\partial T}{\partial r_h} = 0,\,\frac{\partial^2 T}{\partial {r_h}^2} = 0$). Since $R_N$ has only finitely many isolated divergence points in the domain $(r_m,\,\infty)$, this is equivalent to stating that, in the limit $r\to r_h$ from both sides, the solution set of $\lim_{r \to r_h} |R_N| = +\infty$ coincides with the solution set of $T'(r_h)=0$, or
\begin{align}
\left\{ r_h > r_m \,\middle|\, \lim_{r \to r_h} |R_N| = +\infty \right\} = \left\{ r_h > r_m \,\middle|\, T'(r_h) = 0 \right\}.
\end{align}

This also explains why the divergence of the curvature scalar is only a necessary condition for a first-order phase transition, rather than a sufficient one. For example, $A1^+$ and $A1^-$ class black holes have a temperature function with one extremal point, which would lead to a divergence of the curvature scalar, yet no first-order phase transition occurs in these black holes.

\subsection{Example: RN-AdS black hole}
We now take the RN-AdS black hole as a concrete example to rigorously prove that in the RN-AdS black hole, the points at which $R_N$ diverges can only be extremal points or stationary inflection points of the temperature function.

The temperature of the RN-AdS black hole is
\begin{align}
T =\frac{1}{4\pi} \left(\frac{1}{r_h} - \frac{Q^2}{r_h^3} + \frac{3r_h}{\ell^2}\right).
\end{align}
We introduce the dimensionless rescaling
\begin{align}
\tilde{T} = \frac{T}{T_c}, \quad \tilde{V} = \frac{V}{V_c}, \quad \tilde{P} = \frac{P}{P_c}, \quad \tilde{r}_h = \frac{r_h}{r_c},
\end{align}
at the critical point, where
\begin{align}
T_c = \frac{1}{3\sqrt{6}\pi Q}, \quad V_c = \frac{4}{3}\pi r_c^3, \quad r_c = \sqrt{6}Q, \quad P_c = \frac{1}{96\pi Q^2}.
\end{align}
Then the normalized Ruppeiner curvature scalar Eq.~(\ref{RN}) can be cast into
\begin{align}
R_N = \frac{\left( 3\tilde{V}^{\frac{2}{3}} - 1 \right) \left( 3\tilde{V}^{\frac{2}{3}} - 4\tilde{T}\tilde{V} - 1 \right)}{2\left( 3\tilde{V}^{\frac{2}{3}} - 2\tilde{T}\tilde{V} - 1 \right)^2}.
\end{align}
For $R_N$ to diverge, we require
\begin{align}
 3\tilde{V}^{\frac{2}{3}} - 2\tilde{T}\tilde{V} - 1=0.\label{TV0}
\end{align}
We now analyze this equation. Considering
\begin{align}
\tilde{V} = \frac{r_h^3}{6\sqrt{6}Q^3}, \quad \tilde{T} = \frac{3\sqrt{6}Q}{4} \left( \frac{1}{r_h} - \frac{Q^2}{r_h^3} + \frac{3r_h}{\ell^2} \right),
\end{align}
we have
\begin{align}
\tilde{T}\tilde{V} = \frac{r_h^3}{8Q^2} \left( \frac{1}{r_h} - \frac{Q^2}{r_h^3} + \frac{3r_h}{\ell^2} \right),
\end{align}
and
\begin{align}
3\tilde{V}^{\frac{2}{3}} - 2\tilde{T}\tilde{V} - 1 = \frac{1}{4Q^2}\left(r_h^2 - 3Q^2 - \frac{3r_h^4}{\ell^2}\right).
\end{align}
That is, for Eq.~(\ref{TV0}) to hold, we require
\begin{align}
-r_h^2 + 3Q^2 + \frac{3r_h^4}{\ell^2} = 0.
\end{align}
This is consistent with $T'(r_h) = 0$ given by the local criterion (the dimensionless rescaling does not affect the behavior of the function itself). This equation can have either two positive real roots or no positive real roots, corresponding respectively to $A2$ class black holes (with a first-order phase transition) and $B^+$ class black holes (without a first-order phase transition) \cite{Zhang:2025inl}.

To visualize clearly the exact correspondence between the divergence points of $R_N(r_h)$ and the extremal points and stationary inflection points of $T(r_h)$, we plot the dimensionless $R_N(\tilde{r}_h)$ and $\tilde{T}(\tilde{r}_h)$ for the same parameters for comparison. The dimensionless temperature can be written as
\begin{align}
\tilde{T} = \frac{3\tilde{V}^{\frac{1}{3}}}{8} \left( \tilde{P} + \frac{2}{\tilde{V}^{\frac{2}{3}}} - \frac{1}{3\tilde{V}^{\frac{4}{3}}} \right).
\end{align}

\begin{figure*}
	\begin{minipage}{1\hsize}
		\begin{center}
			%\vspace*{10mm}
			
			\subfigure[]{
				\label{rnrh}
				\includegraphics*[scale=0.25]{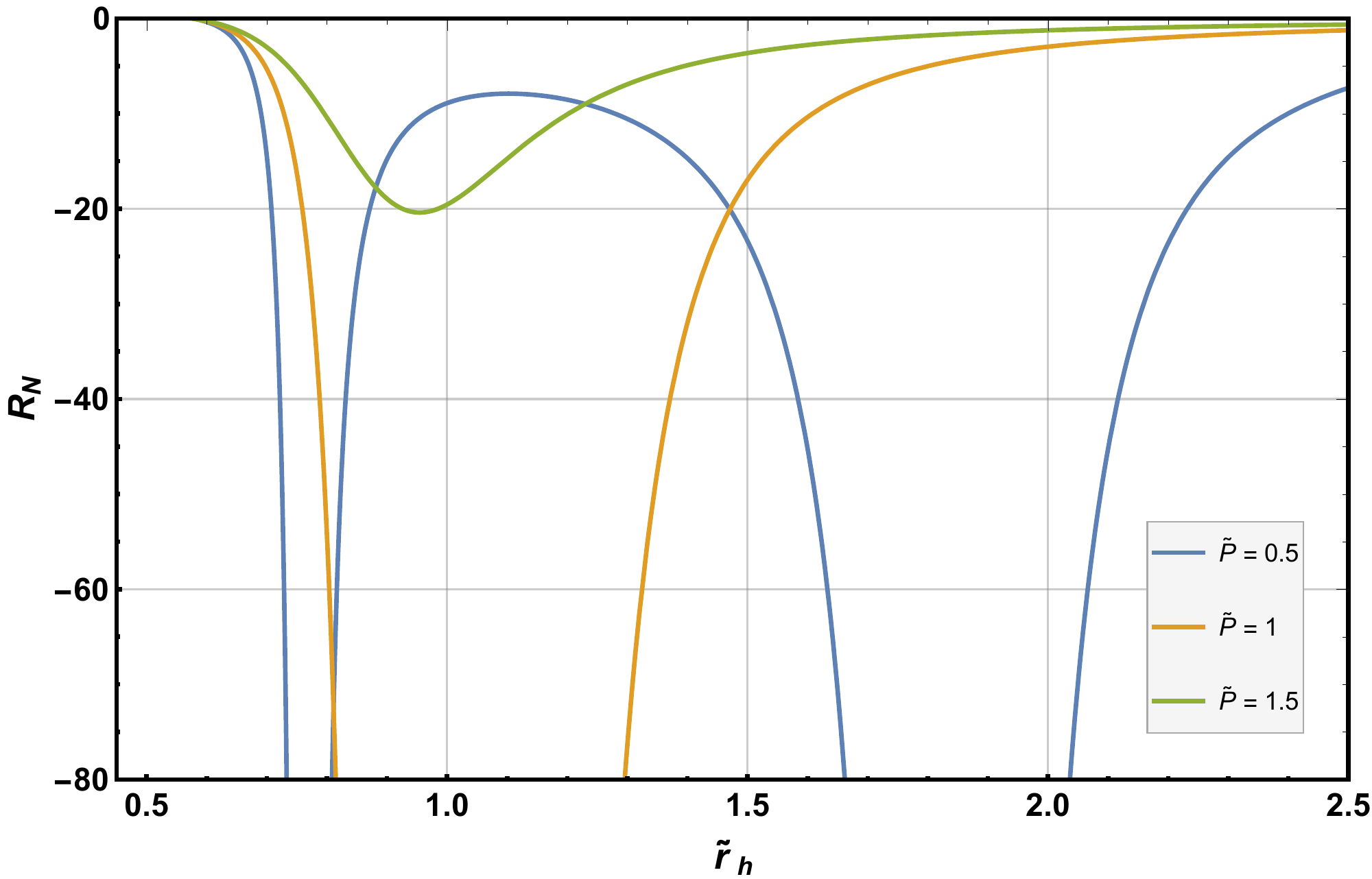}}
			\subfigure[]{
				\label{trh}
				\includegraphics*[scale=0.25]{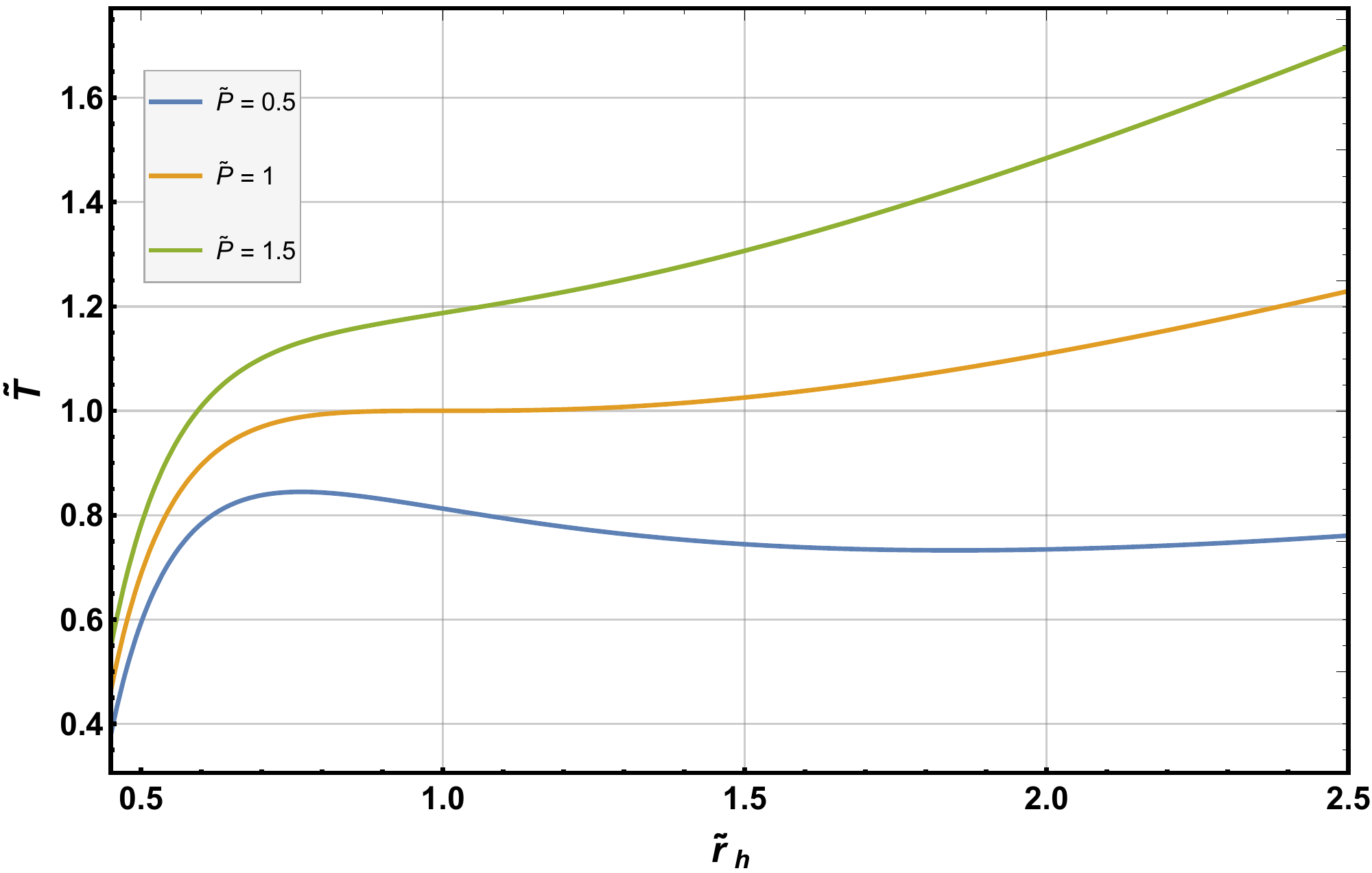}}
            
		\end{center}
		\caption{Comparison between thermodynamic geometry and the local geometric framework for the RN-AdS black hole. (a) The normalized Ruppeiner curvature scalar $R_N$ and (b) the dimensionless temperature $\tilde{T}$ as functions of the dimensionless horizon radius $\tilde{r}_h$. In each plot, three curves are plotted for $\tilde{P}=0.5,1,1.5$, corresponding respectively to the cases with a first-order phase transition, at the thermodynamic critical point, and without a phase transition. The divergence points of $R_N$ in (a) coincide exactly with the extremal points or stationary inflection points of $\tilde{T}$ in (b).
        }
		\label{Fig.1}
	\end{minipage}
\end{figure*}

Fig.~\ref{Fig.1} shows the behavior of $R_N(\tilde{r}_h)$ and $\tilde{T}(\tilde{r}_h)$ for three sets of the same parameters ($\tilde{P}=0.5,1,1.5$, corresponding respectively to the cases with a first-order phase transition, at the thermodynamic critical point, and without a first-order phase transition). It can be seen that at the extremal points and inflection points of $\tilde{T}(\tilde{r}_h)$, $R_N(\tilde{r}_h)$ exhibits divergence consistent with the expected behavior.

\subsection{Weinhold geometry and the local geometric framework}
Unlike the Ruppeiner metric $ds_{(R)}^2$, the Weinhold metric $ds_{(W)}^2$ is constructed from the Hessian matrix of the internal energy. These two metrics differ by a temperature-dependent conformal factor \cite{10.1063/1.446467}:
\begin{align}
ds_{(R)}^2 = \frac{1}{T} ds_{(W)}^2.
\end{align}
For non-extremal black holes, $T>0$, and $T$ is a smooth function of $r_h$. Therefore the conformal factor $1/T$ is smooth and everywhere nonvanishing over the entire domain. This conformal relation provides a physical motivation for expecting that the two geometries share the same singularity structure. However, to establish this rigorously, we directly compare the normalized curvature scalars.

For a general thermodynamic system with first law
\begin{align}
dM = T\,dS + \sum_i Y_i\,dX^i,
\end{align}
where $X^i$ denote the extensive variables and $Y_i$ the conjugate intensive variables. In black hole systems, this term can be expressed as the change in black hole mass caused by generalized forces. In $(T,X)$ coordinates, the normalized Weinhold scalar curvature is given by \cite{Guo:2021wcf}
{\small
\begin{align}
\mathcal{R} = \frac{(1-\Pi) \left[ (\partial_X Y)\partial_{T,X} Y - T(\partial_{T,X} Y)^2 + 2T(\partial_X Y)(\partial_{T,T,X} Y) \right]}{2(\partial_X Y)^2}.
\end{align}}
With $X=V,\,Y=P$, the denominator becomes $2(\partial_V P)^2$, independent of $\Pi$. For the standard AdS black holes considered in this work, the nonlinear electrodynamics (NLED) correction vanishes, $\Pi=0$, and the normalized Weinhold scalar curvature reduces to
{\small
\begin{align}
\mathcal{R}_{N} = \frac{(\partial_V P)(\partial_{T,V} P) - T(\partial_{T,V} P)^2 + 2T^2(\partial_V P)(\partial_{T,T,V} P)}{2(\partial_V P)^2}.
\end{align}}
For regular black holes with NLED terms, $\Pi$ is a finite smooth parameter with $0<\Pi<1$; it appears only in the numerator and thus does not affect the denominator. Since $\mathcal{R}_{N}$ and $R_N$ share the same denominator $2(\partial_V P)^2$, they diverge at the same points. Hence, the conclusion holds for both standard and NLED corrected AdS black holes.

This provides a rigorous proof that the Weinhold geometry inherits the singularity points of the Ruppeiner geometry.

\section{The unified framework for black hole first-order phase transitions}\label{chap:5}
Black hole first-order phase transitions have been extensively studied for over a decade \cite{Mann:2025xrb}. In Ref.~\cite{Zhang:2025inl}, we elevated to a rigorous criterion the phenomenological observation found in numerous studies that the $T(r_h)$ curve possesses two extremal points during a first-order phase transition, and established the local geometric framework, which takes the analysis of the extremal points of the temperature function as its basic method and classification scheme. In subsequent studies, we found that the local geometric framework has exact mappings with the global topological framework and the complex analysis framework \cite{Zhang:2026sht}, and that these exact mappings essentially originate from the geometric structure they share for black hole first-order phase transitions: the catastrophe manifold of a cusp catastrophe \cite{Zhang:2025inl,Zhang:2026catastrophe}. This reveals that the different characteristic information obtained by studying first-order phase transitions with different frameworks is essentially different projections of the cusp catastrophe in different frameworks.

Based on this conclusion, a reasonable and important corollary is that all research frameworks for black hole first-order phase transitions should have mutual mapping relations. In this work, we prove that another important framework in the analysis of black hole first-order phase transitions, namely the Ruppeiner geometry and Weinhold geometry in thermodynamic geometry, indeed has exact mappings with the local geometric framework. Although in this work we have not established a dictionary between the curvature scalar and the topological number $W$ (the core characteristic quantity of the global topological framework) or the Riemann surface foliation number (the core characteristic quantity of the complex analysis framework), through the local geometric framework as an intermediary, the correspondence between $R_N$ and the topological number and the Riemann surface foliation number is straightforward, but the explicit construction of these mappings will be left for future work. With the local geometric framework as the central framework, the four frameworks form a unified framework in the real domain, which will substantially advance our understanding of the mathematical structure behind black hole first-order phase transitions and unify the previously independent characteristic behaviors in various frameworks into a single mathematical framework. This also indicates that the study of black hole first-order phase transitions has entered a stage of rigorous formalization, and several important research frameworks have established solid connections. The complete correspondence table is shown in Table~\ref{st}.

\begin{table*}[htbp]
\centering
\setlength{\tabcolsep}{3.2pt}
\begin{tabular*}{\textwidth}{c@{\extracolsep{\fill}}cccc}
\hline\hline
Local class & 1st order & Global class & $N_{div}$  & Riemann surface\\
\midrule
$A2$     & Yes & $W^{1+}$ & 2 & 3 foliations\\
\midrule
$A1^{+}$  & No  & $W^{0+}$ & 1    & 2 foliations\\
\midrule
$A1^{-}$  & No  & $W^{0-}$ & 1    & 2 foliations\\
\midrule
$B^{+}$   & No & $W^{1+}$ & 0     & 1 foliation\\
\midrule
$B^{-}$   & No & $W^{1-}$ & 0     & 1 foliation\\
\bottomrule
\end{tabular*}
\caption{Classification of black holes in the unified framework. For each local geometric class, we list whether a first-order phase transition occurs, the corresponding global topological class, the number of divergence points of the normalized scalar curvature $N_{div}$, and the number of Riemann surface foliations.}
\label{st}
\end{table*}

\begin{figure*}
	\begin{minipage}{1\hsize}
		\begin{center}
			%\vspace*{10mm}
			
			\subfigure[]{
				\label{3}
				\raisebox{0.88mm}{\includegraphics*[scale=0.4]{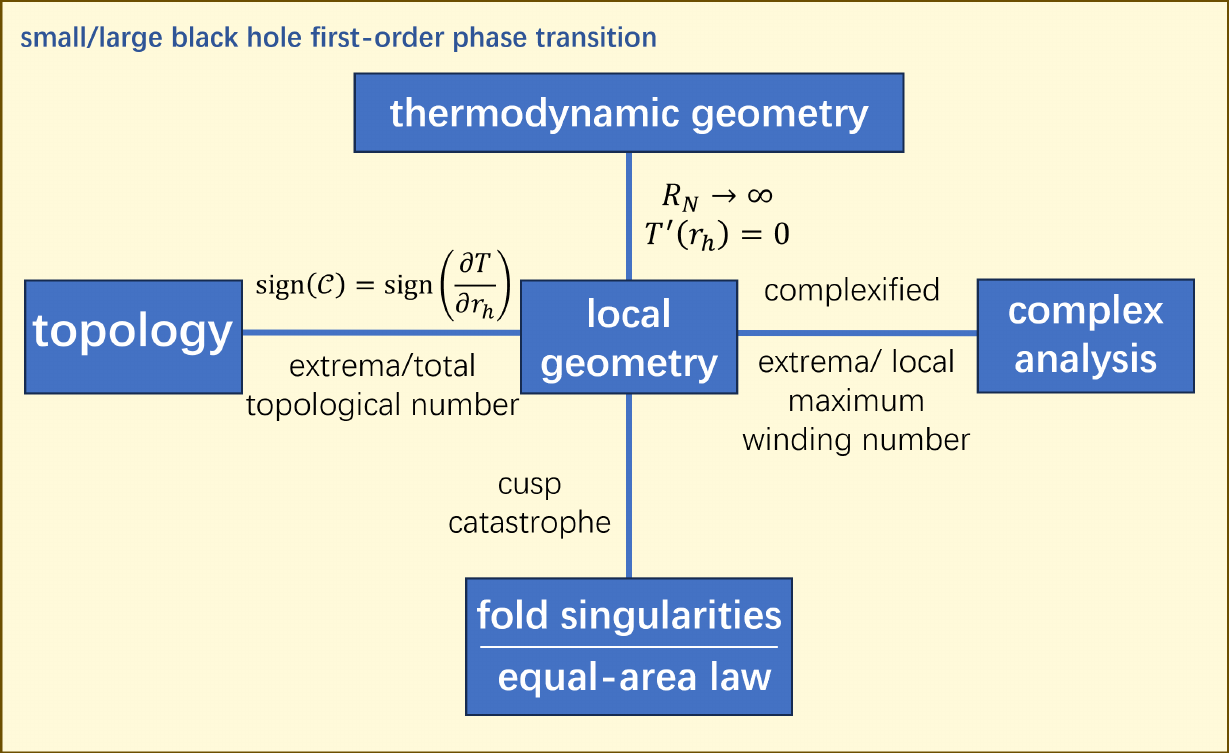}}}
			\subfigure[]{
				\label{4}
				\includegraphics*[scale=0.4]{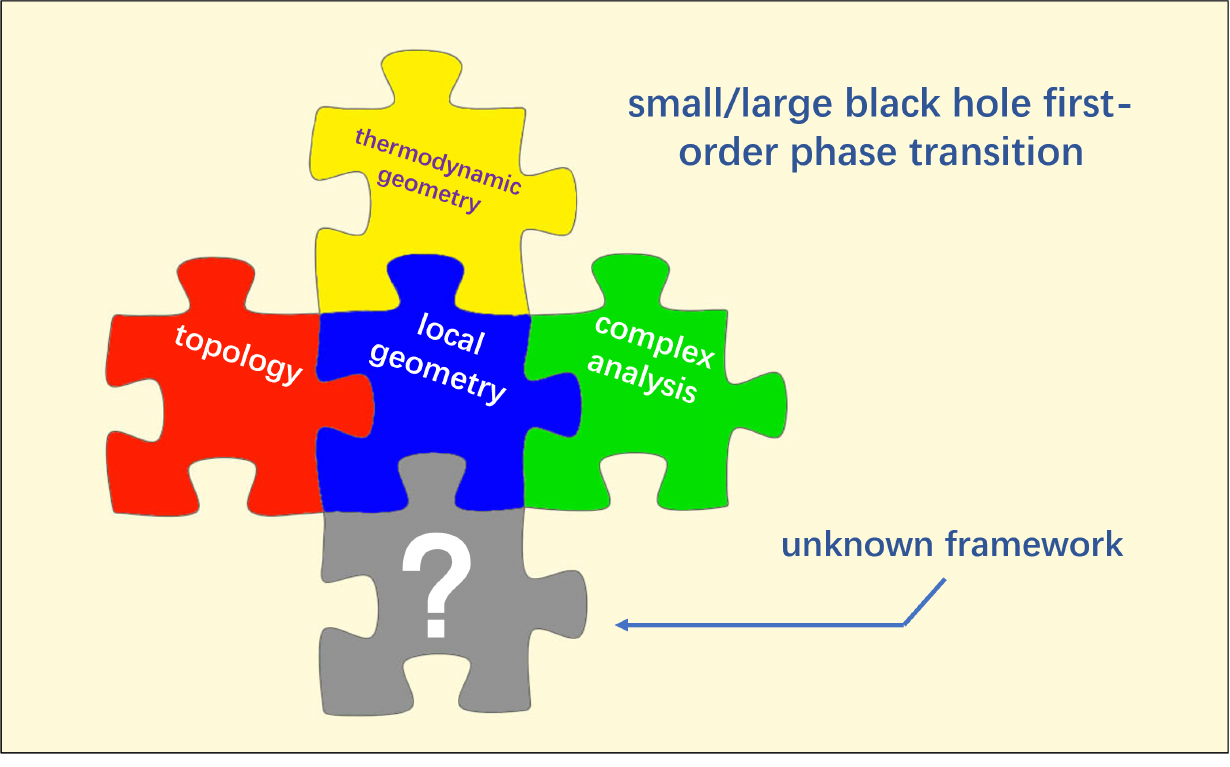}}
            
		\end{center}
		\caption{Unified framework for black hole first-order phase transitions. (a) Rigorous connections among the four frameworks. (b) Schematic illustration of the relations among the four frameworks, presented in the form of a puzzle. Here $\mathcal{C}$ denotes the heat capacity.
        }
		\label{Fig.2}
	\end{minipage}
\end{figure*}

Fig.~\ref{3} shows the rigorous connections among the four unified frameworks at present. Fig.~\ref{4} is a schematic diagram intended to convey the unified relationship among the four frameworks in an intuitive way, but it does not represent the final shape or number of connections that each framework may form in future exploration. As indicated by the gray puzzle piece (representing undiscovered frameworks) in Fig.~\ref{4}, although the four existing puzzle pieces (frameworks) have already been connected through the local geometric framework, we still cannot assert that this picture is complete. Therefore, future research can continue to expand this puzzle on this basis.

\section{Further comments and conclusions}\label{chap:6}
In this paper, by analyzing the behavior of the Ruppeiner curvature scalar $R_N(r_h)$ and the temperature function $T(r_h)$, we find that the set of divergence points of $R_N(r_h)$ coincides exactly with the solution set of $T'(r_h)=0$, i.e., the set of extremal points of $T(r_h)$ (dimensionless rescaling does not change the behavior of the function). This exact correspondence provides a rigorous foundation for the previous approach of using the divergence of $R_N$ to diagnose phase transitions: since $T'(r_h)=0$ is a necessary condition for a first-order phase transition, the divergence of $R_N$ is reliable as a criterion in the sense of screening. On the other hand, because it cannot distinguish between extremal points and inflection points, it cannot alone serve as a sufficient condition for a first-order phase transition. The same analysis also applies to Weinhold geometry.

The successful establishment of the above equivalence relation means that in black hole first-order phase transitions, four important analytical frameworks, namely global topology, thermodynamic geometry, complex analysis, and the local geometric framework, have been unified within a single theoretical framework. Since all four frameworks point to the extremal point structure of the temperature function, using the local geometric framework as the connecting center, one can read off the information of the other frameworks through the corresponding dictionaries. For example, the local geometric framework can identify whether a divergence point of the curvature scalar corresponds to an extremal point or a stationary inflection point by analyzing the number and type of extremal points of the temperature curve, and can simultaneously read off the topological number and the Riemann surface foliation number. A single divergence point of $R_N$ alone cannot distinguish these cases, but when combined with the overall behavior of the temperature curve (the number of extremal points, monotonicity, and whether they merge), it allows precise classification.

This unified framework also opens up several directions for further research. For example, can it be extended to other types of phase transitions, and do there exist undiscovered analytical frameworks that can likewise be reduced to the folding structure of the temperature function?

In summary, the establishment of a unified framework for black hole first-order phase transitions further deepens the understanding of the mathematical structure of first-order phase transitions and provides a generalizable analytical paradigm for analyzing other types of phase transitions.

\section{Acknowledgements}
This work was supported by the National Natural Science Foundation of China (Grant Nos. 12533001, 12473001, 12575049), the National SKA Program of China (Grant Nos. 2022SKA0110200, 2022SKA0110203), the China Manned Space Program (Grant No. CMS-CSST-2025-A02), and the 111 Project (Grant No. B16009).

\bibliography{reference}

\end{document}